\documentclass[preprint,12pt]{elsarticle}

\usepackage{amsmath,amssymb,amsfonts}
\usepackage{multirow}
\usepackage{multicol}
\usepackage{booktabs}
\usepackage{listings}
\usepackage{tablefootnote}
\usepackage[utf8]{inputenc}

\begin{document}

\begin{frontmatter}



\title{Deep learning automates Cobb angle measurement compared with multi-expert observers}


\author[inst1]{Keyu Li \fnref{eqContrib}}
\fntext[eqContrib]{These authors contributed equally to this work.}

\affiliation[inst1]{organization={Department of Electrical and Computer Engineering at Duke University},
            city={Durham},
            postcode={27705}, 
            state={NC},
            country={USA}}

\affiliation[inst2]{organization={Department of Radiology at Duke University},
            city={Durham},
            postcode={27705}, 
            state={NC},
            country={USA}}
\affiliation[inst3]{organization={Department of Orthopaedics at Duke University},
            city={Durham},
            postcode={27705}, 
            state={NC},
            country={USA}}

\affiliation[inst4]{organization={Department of Pediatric Radiology at Duke University},
            city={Durham},
            postcode={27705}, 
            state={NC},
            country={USA}}
\affiliation[inst5]{organization={Department of Computer Science at Duke University},
            city={Durham},
            postcode={27705}, 
            state={NC},
            country={USA}}
            
\author[inst1]{Hanxue Gu \fnref{eqContrib}}
\author[inst2]{Roy Colglazier}
\author[inst2]{Robert Lark}
\author[inst2]{Elizabeth Hubbard}
\author[inst2]{Robert French}
\author[inst3]{Denise Smith}
\author[inst1]{Jikai Zhang}
\author[inst2]{Erin McCrum}
\author[inst3]{Anthony Catanzano}
\author[inst4]{Joseph Cao}
\author[inst2]{Leah Waldman}
\author[inst1,inst2,inst5]{Maciej A. Mazurowski}
\author[inst3]{Benjamin Alman}

\begin{abstract}
Scoliosis, a prevalent condition characterized by abnormal spinal curvature leading to deformity, requires precise assessment methods for effective diagnosis and management. The Cobb angle is a widely used scoliosis quantification method that measures the degree of curvature between the tilted vertebrae. Yet, manual measuring of Cobb angles is time-consuming and labor-intensive, fraught with significant interobserver and intraobserver variability. To address these challenges and the lack of interpretability found in certain existing automated methods, we have created fully automated software that not only precisely measures the Cobb angle but also provides clear visualizations of these measurements. This software integrates a deep neural network-based spine region detection and segmentation, spine centerline identification, pinpointing the most significantly tilted vertebrae, and direct visualization of Cobb angles on the original images.
Upon comparison with the assessments of 7 expert readers, our algorithm exhibited a mean deviation in Cobb angle measurements of 4.17 degrees, notably surpassing the manual approach's average intra-reader discrepancy of 5.16 degrees. The algorithm also achieved intra-class correlation coefficients (ICC) exceeding 0.96 and Pearson correlation coefficients above 0.944, reflecting robust agreement with expert assessments and superior measurement reliability.
Through the comprehensive reader study and statistical analysis, we believe this algorithm not only ensures a higher consensus with expert readers but also enhances interpretability and reproducibility during assessments. It holds significant promise for clinical application, potentially aiding physicians in more accurate scoliosis assessment and diagnosis, thereby improving patient care. 

\end{abstract}

\begin{graphicalabstract}
\includegraphics[width=\textwidth]{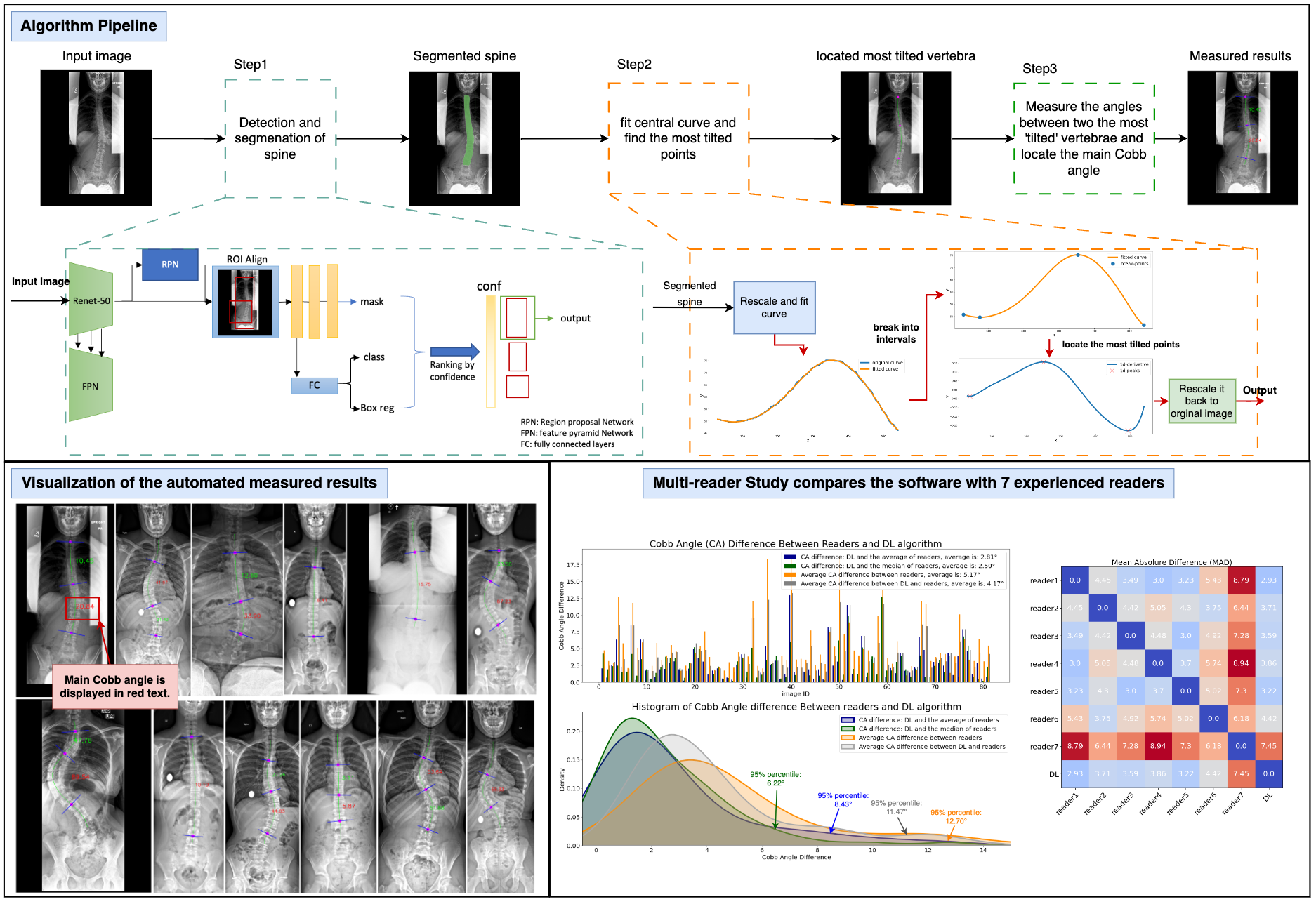}
\end{graphicalabstract}


\end{frontmatter}


\section{Introduction}
\label{sec:intro}
Idiopathic scoliosis is a three-dimensional spinal deformity, defined by an abnormal curvature of more than 10 degrees on a coronal radiograph with no clear evidence for other underlying diseases based on the history and findings \cite{yaman2014idiopathic,weinstein2008adolescent}. It is the most common type of scoliosis and is most often encountered by primary care physicians, pediatricians, and spinal surgeons in the adolescent population, designated adolescent idiopathic scoliosis (AIS) \cite{weinstein2008adolescent}. Early and accurate diagnosis plays a significant role in choosing the treatment type and making the treatment plan. When diagnosing scoliosis, physicians rely on the severity and magnitude of the spine deformity and the risk of progression when deciding the best management method \cite{weinstein2008adolescent, altaf2013adolescent}. In clinical evaluations, a physical exam is often the provider’s first method of detecting abnormal spinal curves. The Adam’s forward bend test or use of a scoliometer allows the examiner to determine if further tests or imaging are warranted and helps to avoid over-radiating children without grossly abnormal spinal curves \cite{devito2010clinical,horne2014adolescent}.

When physical examination of the patient is concerned for abnormally curved spines,  posteroanterior and lateral radiographs of the spine are the most common diagnostic modality in AIS. The Cobb angle \cite{yaman2014idiopathic}, is the standard of quantification for the diagnosis and analysis of scoliosis, measured as the angle formed by the intersection of two lines that are parallel to the direction of the most tilted vertebrae. A patient with a 10° or greater Cobb angle is considered to have scoliosis \cite{kim2010scoliosis}. In addition to growth potential, Cobb angle is the most important factor when predicting the risk of spinal curve progression past skeletal maturity. 
 
Traditional methods of measuring the Cobb angle require experts’ manual annotation on radiographs, which is tedious and time-consuming. Also, due to large abnormal variation among different patients \cite{wu2017automatic} and measurement errors caused by vertebral rotation, the position of patients \cite{kim2010scoliosis}, and low image quality, physicians’ manual measurements suffer large inter-reader and intra-reader variability \cite{zhang2010computer}. 

In this study, we proposed a robust automated Cobb angle measurement method using deep learning. Instead of segmenting each individual vertebra \cite{wu2017automatic, horng2019cobb, alharbi2020deep, fu2021automated, khanal2020automatic, yi2020vertebra, huang2024comparison, reformat2023validation, suri2023conquering}, we considered the geometric properties of the entire spine based on the instance-level segmentation and a central line fitting. All Cobb angles are measured using tolerance-based derivatives of the centerline after the fitting process. 

In the experiment, we applied rich statistical analysis, such as inter-rater reliability, Cohen’s kappa, and the Pearson correlation coefficient to evaluate our model's performance. We experimentally compared the measurement generated by our algorithm with measurements of multiple physicians and showed that our model performs at the level of a human expert. Furthermore, taking into account the substantial costs associated with expert annotation in actual clinical environments, our model distinguishes itself from numerous prior methodologies. Notably, it neither necessitates human expert annotations for its training nor relies on individual vertebrae for angle measurements.

Our contributions are the following:
\begin{enumerate}
    \item A powerful, completely autonomous, deep learning-based algorithm for measuring Cobb angles with low annotation costs.
    \item An innovative and reliable approach for measuring the Cobb angle with tolerance-based derivatives that consider the morphology of the whole spine rather than a single vertebra and is compatible with clinical practice.
    \item Extensive visibility of the Cobb angle measuring procedure and results, which is helpful in clinical settings.
    \item Comprehensive statistical analysis of model performance compared to multiple readers and precise Cobb angle measurement equivalent to human specialists.
\end{enumerate}

\section{Related Work}
In recent years, several attempts have been made to develop computer-aided methods for measuring Cobb angles for 2D images, which can be divided into two categories: image-enhancement-based methods and machine-learning-based methods \cite{jin2022review}.
 
Although the image enhancement methods work well on some spine images, they come with a very high computational cost and a lack of robustness on different images since they require precise feature engineering. In contrast, lots of machine learning methods can extract image features automatically and have relatively stable performance among images. Most machine learning methods \cite{jin2022review, kh2021automatic} mainly focus on segmenting individual vertebra \cite{horng2019cobb, reformat2023validation, caesarendra2022automated, zhao2022automatic} or predicting vertebral landmarks \cite{wu2017automatic, alharbi2020deep, fu2021automated, khanal2020automatic, huang2024comparison, suri2023conquering, sun2022comparison, qiu2023mma}. However, due to the shape variation of each vertebra, those methods that are heavily based on individual vertebrae are not accurate enough \cite{kh2021automatic}, which can lead to deviation of final Cobb angles. Also, many previous machine-learning-based methods needed a manual selection of upper and lower vertebrae, which introduced subjectivity.
 
Instead of using individual vertebrae or landmarks to measure the Cobb angle, some researchers conduct Cobb angle estimation by using the whole spine's curvature. Tu, Y. et al. \cite{tu2019automatic} proposed a model that calculated the Cobb angle based on the segmentation of the whole spine. However, their research was limited by low robustness due to a small test set as well as a lack of enough statistical evaluation methods. Okashi et al. \cite{al2017automatic} directly used the centerline of the spine to calculate Cobb angles. However, their method involved a complicated image processing algorithm while requiring tedious labor. Dubost, et al. \cite{dubost2020automated} developed a cascaded network to segment the centerline to measure the Cobb angle. In their algorithm, two cascaded convolutional neural networks were used first to segment the complete spine and then segment the centerline based on the whole spine mask. Compared to our method, adopting the second network to obtain the centerline introduces extra computational cost, which is not ideal considering the scarce computing resources in clinical scenarios.  Similar to our method, Zhou et al. \cite{zhou2023vertebral} and Bernstein et al. \cite{bernstein2021radiographic} focused on extracting vertebral center points for Cobb angle measurement with low annotation cost. However, both methods lack sufficient robustness due to the limited number of vertebral center points predicted by their models, while the performance of \cite{zhou2023vertebral} was constrained by a small test set size and a limited number of surveyors (i.e., only two surveyors), as indicated by intra-class correlation coefficients (ICC).


In our study, we present a deep-learning-based pipeline offering several advantages: 1) Focusing on the entire spine rather than individual vertebra, our method minimizes annotation and computational costs, while also eliminating the need for domain experts. 2) Our model efficiently trains end-to-end while providing fully automated Cobb angle measurements in clinical settings without manual intervention. 3) Employing a tolerance-based mechanism, our model achieves high consensus with multi-expert readers while also ensuring robustness and reliability even in low-quality X-rays, which are demonstrated in our result section. Note that our model does not have any requirement for any special hardware or software.

\section{Materials and Methods}
\label{sec: method}
\begin{figure}
    \centering
    \includegraphics[width=\textwidth]{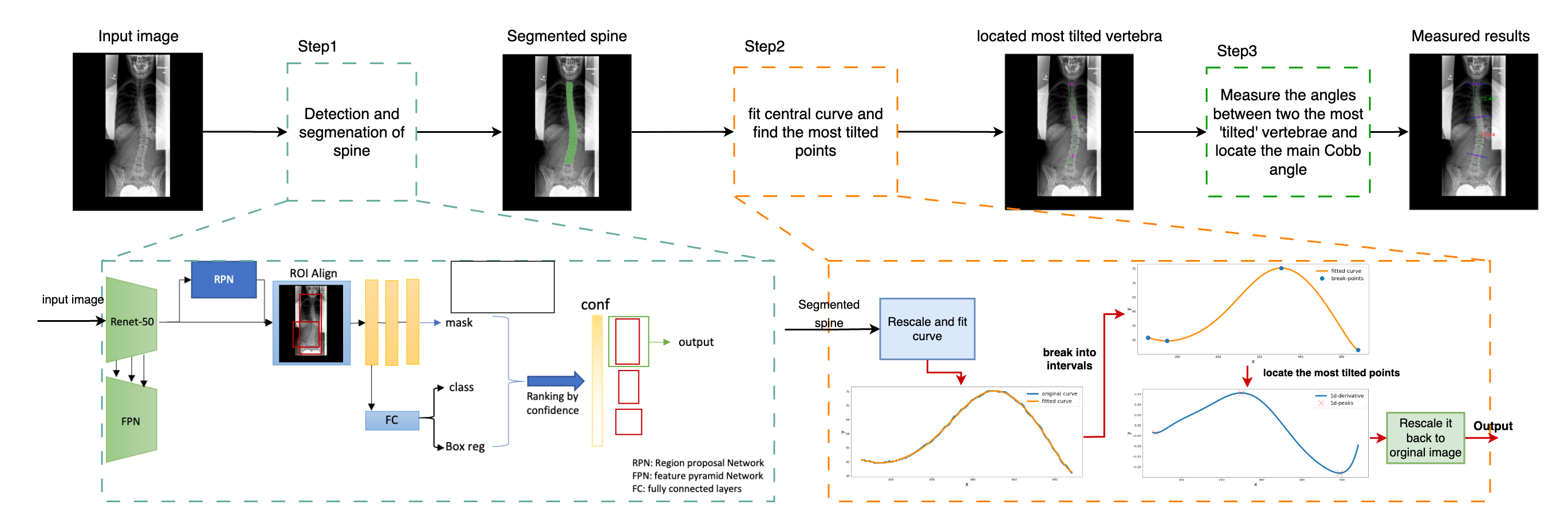}
    \caption{Visualization of our proposed pipeline. Our approach consists of three steps: 1) the identification of ROI and segmentation of the spine; 2) Fit the spine's central curve and identify the vertebrae that are most 'tilted' 3) Calculate the Cobb angle; all Cobb angles are presented on the image, with the main Cobb angle indicated in red.}
    \label{fig:pipeline}
\end{figure}
\subsection{Data Preparation}
\begin{table}[]
\caption{Statistics of the dataset, plus-minus values are means+- std.}
\label{Tab: data}
\fontsize{9pt}{9pt}\selectfont
\begin{tabular}{llll}
\hline
Group                   & train \&   validation group (N=830) & \begin{tabular}[c]{@{}l@{}}test group\\    \\ (N=81)\end{tabular} & total (N=921) \\ \hline
Age-yr.                 & 12.89$\pm$3.81                         & 12.02$\pm$3.39                                                       & 12.67$\pm$3.52   \\
Female sex - no   (\%)  & 501 (60.4\%)                        & 51 (63.0\%)                                                       & 552(59.9\%)   \\
Race - no (\%)          &                                     &                                                                   &               \\
-black                   & 192 (23.1\%)                        & 19(23\%)                                                          & 211           \\
-Caucasian/white         & 494 (59.5\%)                        & 46(56.8\%)                                                        & 540           \\
-Asian                   & 30 (3.6\%)                          & 1(1.2\%)                                                          & 31            \\
-other                   & 58 (7.0\%)                          & 9(11.1\%)                                                         & 67            \\
-not reported/declined   & 56(6.7\%)                           & 6(7.4\%)                                                          & 62            \\
Ethnic group - no. (\%) &                                     &                                                                   &               \\
-Hispanic or Latino      & 69(8.3\%)                           & 7(8.6\%)                                                          & 76            \\
-not Hispanic or Latino  & 705(84.9\%)                         & 70(86.4\%)                                                        & 775           \\
-not reported/declined   & 56(6.7\%)                           & 4(4.9\%)                                                          & 60            \\ \hline
\end{tabular}
\end{table}

This study was approved by the Duke University Health System (DUHS) institutional review board. Because of the retrospective nature of the study, informed consent was waived. Specifically, in the Duke Electronic Medical Record Database, we identified spine X-ray and scoliosis X-ray imaging studies at Duke from January 1, 2014, to Nov 3rd year, 2020, and excluded studies for patients who were 18 years of age or older. From this filtered dataset, we took a convenience sample as our analytical dataset, consisting of 1084 patients and 2294 studies. Then, we queried the Picture Archiving and Communication System (PACS) server at DUHS to retrieve the images for all studies in the analytical dataset.

 Next, only images that met all the following inclusion/exclusion criteria listed below were retained for model development:
 \begin{enumerate}
     \item Only images with PA/AP spinal view were selected;
     \item Only images with the entire spine were selected;
     \item Only images without spinal hardware were selected;
     \item Only images with no obstruction of the spinal area were selected;
     \item Only images with high image quality were selected.
 \end{enumerate}

\subsection{Training, Validation and Test set}
The inclusion/exclusion criteria resulted in a collection of 1460 images belonging to 1436 studies from 830 patients. These images were first split into training and validation sets; next, several images were removed from the training set to ensure there was no overlap of patients between the training and validation subsets. The training data set resulted in 1405 images belonging to 1383 studies from 810 patients. The validation set resulted in 20 images belonging to 20 studies from 20 patients. From the 20 images in the validation set, 12 were randomly selected for physician-annotated Cobb angles.

As for our test set, we downloaded 200 consecutive pediatric cases from Jan 1st,
2021, which contained 192 patients. After deidentifying all DICOM files and constructing corresponding PNG files, we used the same inclusion/exclusion criteria as the training set to select 81 cases consecutively starting from Jan 1st, 2021, while ensuring the cases belonged to new patients and we only had one image per patient. The test set resulted in a collection of 81 images from 81 patients.
The statistics of the training, validation, and test sets are shown in Table. \ref{Tab: data}.

\subsection{Deep learning-based Cobb angle measurement}
Our method consists of three main steps: (1) spine detection and segmentation; (2) the most tilted vertebrae location; and (3) angle measurement. In the first step, we use a single neural network to detect the ROI of the spine area while simultaneously segmenting the entire spine. To measure the Cobb angle, we first perform the spine curve fitting after sampling on the centerline. Then, we calculate the spine derivative to find the most tilted vertebra. We finally finished measuring the Cobb angles by calculating the angles between the most tilted vertebrae, following the definition \cite{kim2010scoliosis}.

\subsubsection{Spine detection and segmentation}
\label{Sec: Spine detection and segmentation}
Considering spines have a relatively consistent appearance and are present as a single distinguishable object in the radiograph and that the raw radiograph may have a large coverage of the body with a relatively small ROI, we employ the instance segmentation algorithm Mask-RCNN \cite{he2017mask} that can simultaneously detect and segment spines. Mask-RCNN was proposed based on Faster RCNN \cite{girshick2015fast}, adding a mask segmentation branch, so it can achieve both high-quality segmentation and ROI detections using a single model setting. We employed a Resnet-50 with a feature pyramid network (FPN) as the backbone. The final model was selected as the best-performing based on the highest mAP on the evaluation set. During inference, instead of setting a predetermined confidence threshold (like 0.9) to get multiple detected objects, we sort the output boxes by confidence value and select the one with the highest confidence. The illustration of our detection and segmentation algorithm is shown in Figure \ref{fig:pipeline} below.

In this step, we implemented a single model setting as opposed to the more typical detection model plus a semantic segmentation model \cite{yi2020vertebra, al2017automatic, dubost2020automated} used in earlier works for the following reasons. First, a single model is more effective and requires less computing power. Second, the pixel-level semantic segmentation model is less resistant to local disturbances in images, while the instance-level model can better concentrate on the object's overall morphological features and support maintaining the spine's structural integrity.

\subsubsection{Location of the most tilted vertebrae}
Finding the most tilted vertebrae can be difficult, thus we equate it geometrically to find the largest curvature on the spine’s central curve. The central curve of the spine was produced in two steps. First, we located the central point in each row of the segmented spine mask and connected those points to obtain the spine's central line. To assure the continued derivability, we smoothed it using a polynomial curve fitting with a maximum of 10 components. Before curve fitting, all spine centerlines were re-scaled to a length of 572 pixels to ensure the creation of our curve fitting algorithms that are applicable to all scenarios.

To find the largest curvature, we first divided the curve into intervals. We then partitioned our curve by breakpoints under two criteria: 1) If a single vertebra is located at the concave or convex points of the curve, we would divide our curves at this point. 2) If this interval is excessively long without convex/concaves, it indicates that the "tilt" is not severe in this region. Nonetheless, the spine may probably tilt gradually within this region; thus, we divided it in the middle. We then found the most tilted vertebra within each interval. Since the spine curve can be seen as a polynomial function f(x), finding the largest tilt can also be seen as finding the point with the largest 1-order derivative. We transfer this problem to find
\begin{equation}
\max_{x \in [x_i, x_{i+1}]} f'(x),
\end{equation}
where $x_i$ and $x_{i+1}$ are two adjacent breakpoints.

\subsubsection{Cobb Angle Measurements}
\label{sec: Cobb angle measurements}
\begin{figure}
    \centering
    \includegraphics[width=\textwidth]{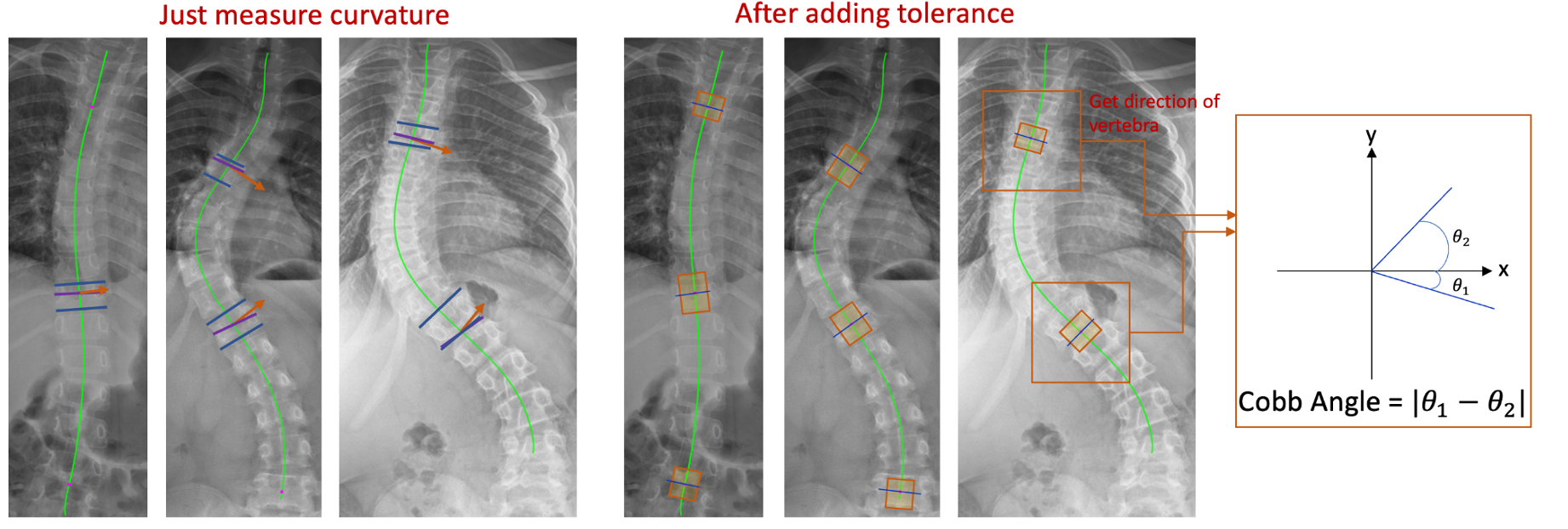}
    \caption{The first three examples demonstrate that there is some offset between the vertebral orientation and the spine's tangent (shown by the orange arrow). After adding the tolerance range to average the curvature, we could measure the Cobb angles.}
    \label{fig:method2}
\end{figure}
To measure the angles between the most "tilted" points, we take the tangent at those points as the direction of vertebrae directly. However, the spine functions more like a chain than a collection of discrete points, with each link altering the mobility of the vertebrae it connects. In addition, in clinical practice, readers typically measure the orientation of the endplates of the vertebrae (the upper and lower blue lines in Figure \ref{fig:method2}), demonstrating that readers may also measure a position that is somewhat off from the geometrically steepest point. Thus, the direction of the steepest point may deviate to some extent from the direction that the vertebrae are measured (the short purple line). After identifying these possible biases, we included a tolerance variable $L_t$ to estimate the average curvatures within a range to determine vertebra orientation instead of utilizing one single point. Our method sets tolerance $L_t$ at 0.15 of each interval length, and the tolerance was set best on the largest agreements with the 8 readers’ measurements on the evaluation set.
\begin{figure}
    \centering
    \includegraphics[width=\textwidth]{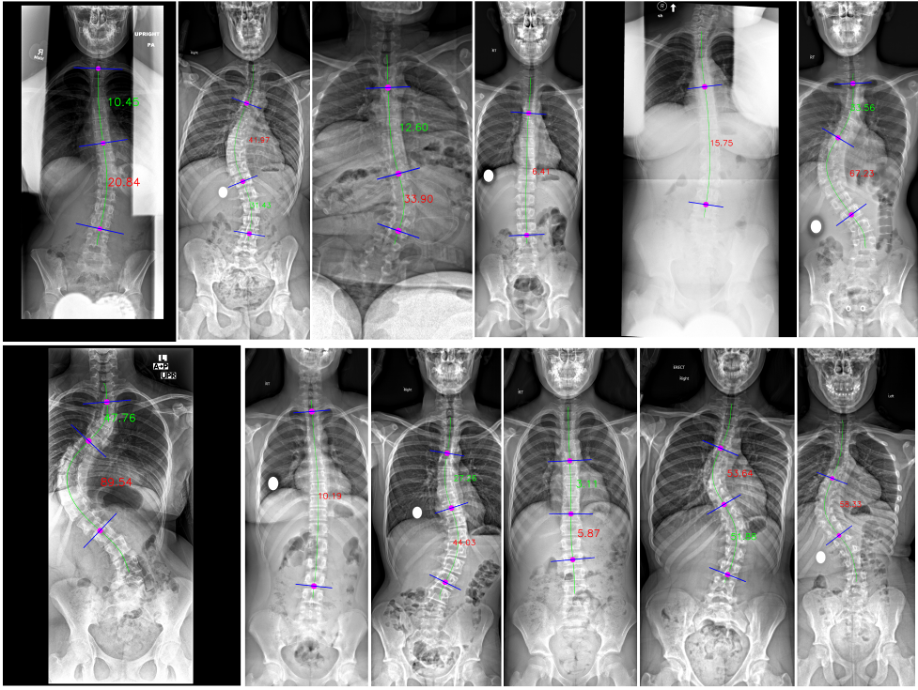}
    \caption{Measured examples of Cobb angles. They are automatically shown as the final outputs of our method, without any further manual sketching. The main Cobb angle is displayed in red lettering, while additional Cobb angles are displayed in green.}
    \label{fig:example_results}
\end{figure}

After measuring vertebral direction $(-90^\circ,90^\circ)$. We then measured the Cobb angles between connected vertebrae. Figure \ref{fig:example_results} illustrates several examples of measured Cobb angles. Our approach depicted the Cobb angles and the vertebrae we referred to, and the main Cobb angle is the one we care most about and the one referred to decides the scoliosis severity \cite{altaf2013adolescent}, colored in red in Figure \ref{fig:example_results}.

\subsection{Evaluation Metrics}
\noindent\textbf{Metrics for spine detection and segmentation algorithm}:
As discussed in Section \ref{Sec: Spine detection and segmentation}, some earlier studies separated the steps of detecting and segmenting. We also implemented a Faster-RCNN + Unet structure as one baseline to be compared. The detection and segmentation algorithms were evaluated on an evaluation set by 1) the mean average precision for detection results; and 2) the Dice Coefficient for segmentation results. 3) the average time it takes to extract the spine from a single x-ray.

\noindent\textbf{Evaluation of Cobb angle measurements}
As mentioned in Section \ref{sec: Cobb angle measurements}, while there are multiple Cobb angles in scoliosis deformity, the analysis focused on the Main Cobb angle, defined as the largest Cobb angle. Thus, to evaluate the algorithm performance, we compared the main Cobb angle of our measurements with 7 readers on the test set under the following metrics: (1) the average absolute difference between the mean and the median measurements of the readers for each case. It is calculated by
\begin{equation}
\begin{aligned}
   \overline{D_{\text {mean }}}&=\frac{1}{n} \sum_{\mathrm{k}=1}^n D_{\text {mean }}=\frac{1}{n} \sum_{\mathrm{k}=1}^n\left|R_{\text {mean }, k}-D_k\right| \\
\overline{D_{\text {median }}}&=\frac{1}{n} \sum_{\mathrm{k}=1}^n D_{\text {median }}=\frac{1}{n} \sum_{\mathrm{k}=1}^n\left|R_{\text {median }, k}-D_k\right| \\ 
\end{aligned}
\end{equation}
, where $R_{mean,k}$ is the mean value of the main Cobb angle measured by 7 readers for case k and $R_{median,k}$ is the median value for that of case k; (2) the average pairwise difference between readers for each class: $\overline{D_{\text {readers }}}=\frac{1}{n} \sum_{\mathrm{k}=1}^n \sum_{i \neq j}^{i, j \in[1, m]} \frac{\left|R_{i, k}-R_{j, k}\right|}{m(m-1)}$, and the average difference between readers and DL algorithm $\overline{D_{\text {readers }-DL}}=\frac{1}{n} \sum_{\mathrm{k}=1}^n \sum_{i=1}^m \frac{\left|D_k-R_{i, k}\right|}{m}$, where $m$ is the number of readers; (3) Mean absolute difference compared with other readers; (4) Pearson correlation coefficient; (5) Intraclass correlation coefficient.

\subsection{Metrics for Scoliosis Severity Classification}
We labeled the different severity levels of scoliosis as 0-4, where Level 0 was Cobb angles $\leq$ 10, level 1 with 10-25, level 2 25-45, level 3 45-60, level 4 $\geq$ 60 and correlated the measured Cobb angles to a multi-classification task for scoliosis severity grading \cite{chowanska2012school}. To evaluate the scoliosis severity grading, we introduced the following metrics: 1) Accuracy and F1-score and 2) Cohen's kappa coefficient.

\section{Evaluation Results}
\begin{figure}
    \centering
    \includegraphics[width=\textwidth]{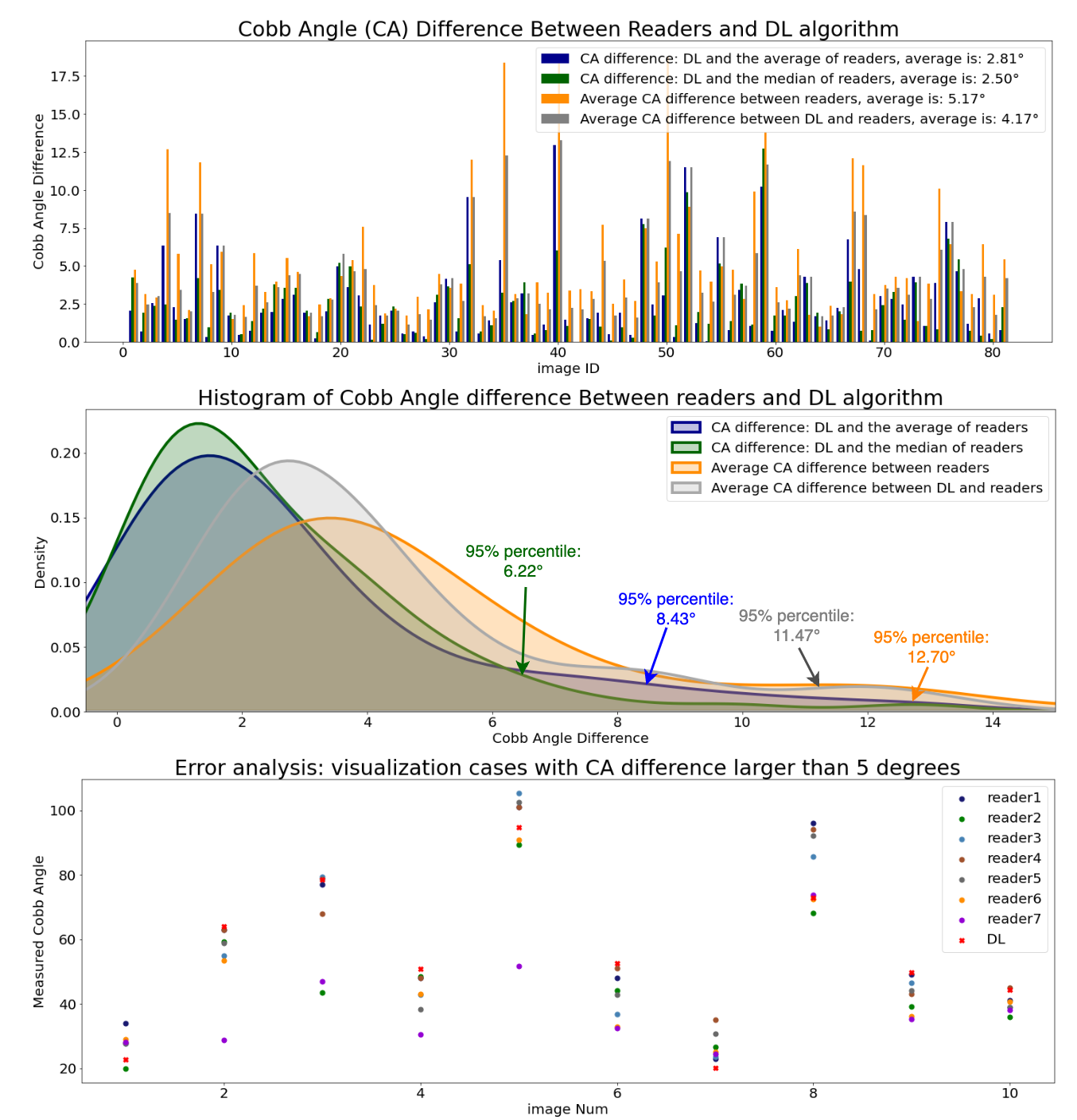}
    \caption{Quantitative evaluation of our algorithm 1) The first column is the measured Cobb angle difference between the average and the median of the readers and our algorithm (blue and green bars), the average pairwise difference among readers (orange bars), and the average difference between readers and our algorithm (gray bars) on the 81 test cases. 2) The second row depicts histograms that elucidate the aforementioned Cobb Angle (CA) discrepancies across the test samples, and 3) Correspondingly, the third row offers a visualization of instances where the CA differential exceeds 5 degrees.}
    \label{fig:quantative_result}
\end{figure}

\subsection{Results on spine detection and segmentation}
\begin{table}[]
\label{tab: seg_results}
\caption{The statistics of the detection and segmentation performance of two spine segmentation pipelines.}
\fontsize{9pt}{9pt}\selectfont
\begin{tabular}{l|lllll}
\hline
Methods          & mAP(IoU=0.5:0.95) & mAP(IoU\textgreater{}0.5) & AR   & DSC   & \begin{tabular}[c]{@{}l@{}}test time (s)\\    \\ (Each image)\end{tabular} \\ \hline
Mask-RCNN       & 0.735             & 0.897                     & 0.79 & 0.917 & 0.63                                                                       \\
Faster-RCNN+Unet & 0.731             & 0.883                     & 0.76 & 0.903 & 1.42                                                                       \\ \hline
\end{tabular}
\end{table}
Compared with the two-model setting Faster-RCNN+Unet results, Mask-RCNN could improve the dice coefficient score of 0.014, shown in Table \ref{tab: seg_results}. Furthermore, it takes less time to infer an image for the Mask-RCNN single-model setting.

\subsection{Results on Cobb angle measurements}
Figure \ref{fig:example_results} provides representative illustrations of the measurements derived from our algorithm. This figure elucidates the extracted spinal centerline, the quantified multi-point Cobb angles, and the orientation of the vertebrae, depicted via a concise blue line. As evidenced by the samples presented, our algorithm demonstrates consistent performance and robust stability across a diverse patient cohort, ranging in scoliosis severity and varying radiographic conditions.

Figure \ref{fig:quantative_result} (top1) displays the Cobb angle distance in the entire test set between our DL algorithm and the two "ground truth" values ($R_{mean}$ and $R_{median}$) those we defined. Our algorithm archives an average absolute distance of $\overline{D_{\text {mean }}}=2.80^{\circ}\left[95 \%\right.$ confidence interval $(\mathrm{CI}): 2.00^{\circ}, 2.99^{\circ}]$ and $\overline{D_{\text {median}}}=2.50^{\circ} [\mathrm{CI}: 2.20^{\circ}, 3.41^{\circ}]$ for all the test cases.  Moreover, upon calculating the Cobb angle (CA) difference between our algorithm and the assessments of the readers, our method yields a mean CA difference of $4.17^{\circ}[\mathrm{CI}: 3.54^{\circ}, 4.79^{\circ}]$, and the CA difference observed amongst the readers themselves are $5.17^{\circ}[\mathrm{CI}: 4.31^{\circ}, 6.04^{\circ}]$. And for 95\% of the test cases, we have a measurement smaller than $6.22^{\circ}$ differences between the median of the readers ($R_{median}$), and $8.43^{\circ}$ difference between the mean of the readers ($R_{mean}$). 

\begin{figure}
    \centering
    \includegraphics[width=\textwidth]{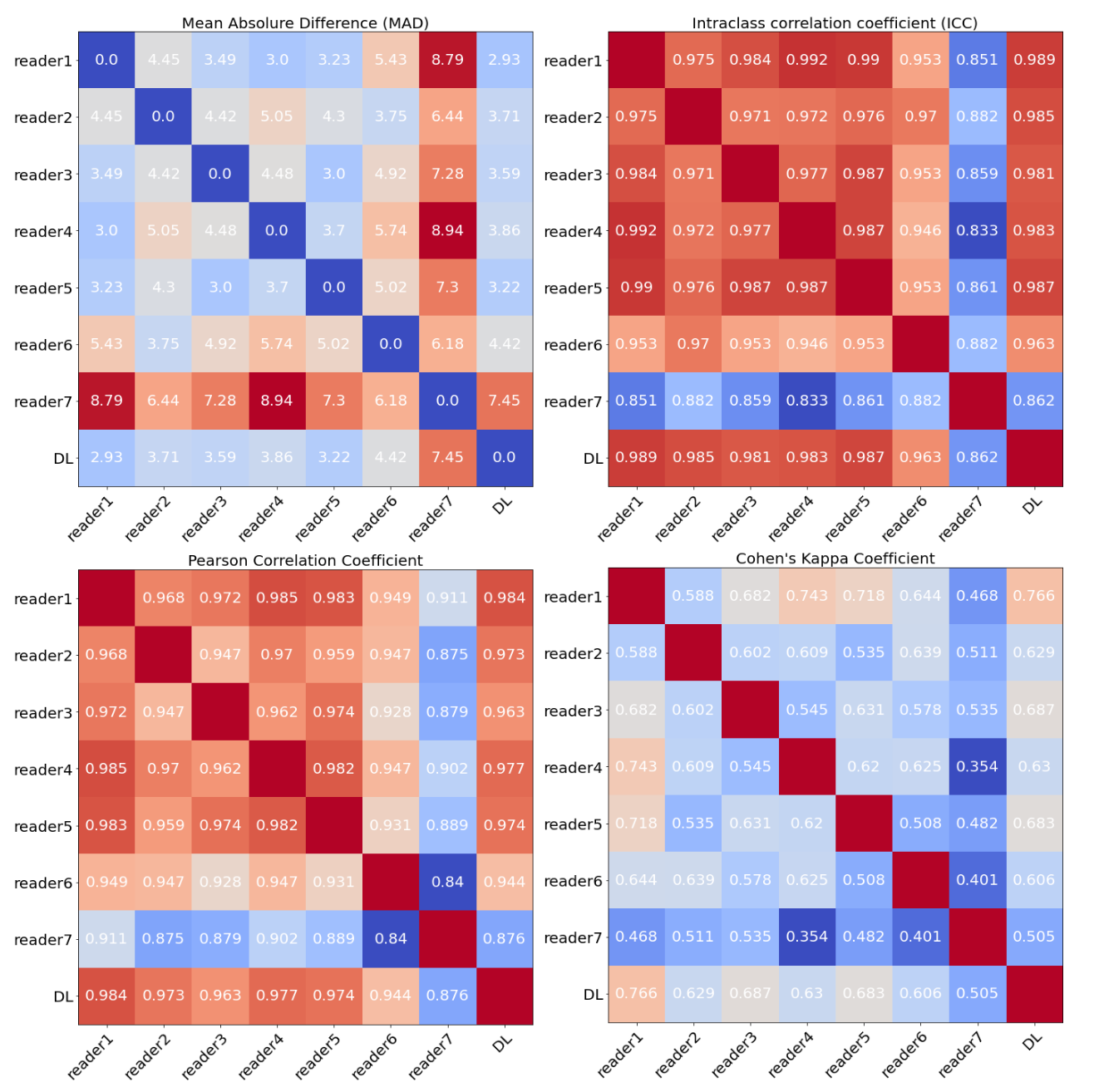}
    \caption{The top left is the pairwise mean absolute difference between the readers and DL, and the top right column is Pearson's coefficient; the bottom left is the intraclass coefficient (ICC) for Cobb angle measurements. The last is the Cohen kappa score for scoliosis severity classification, respectively. }
    \label{fig:MAE_result}
\end{figure}

From the MAD shown in Figure \ref{fig:MAE_result} (row 1, left), we can see that the average pairwise distance between reader 1, reader 2, … and reader 7 to all other readers is 4.73, 4.74, 4.60, 5.15, 4.43, 5.17, and 7.49, respectively, while the average MAD from DL to other readers is 4.17. 
Furthermore, the reliability of the Cobb angle measured by our method was also assessed by the pairwise ICC and Pearson correlation coefficients. Apart from reader 7, our DL algorithm has pair-wise correlation coefficients with all the other readers that are higher than 0.94 and an ICC that is higher than 0.96, proving that it is well-matched to the readers' evaluations.  

\subsection{Results on scoliosis severity classification}
The pairwise Cohen Kappa scores illustrate that our algorithm also has a high level of agreement in grading the severity of scoliosis with the experienced readers, with an average Cohen Kappa of 0.65 with 7 readers. The average accuracy for the multi-class classification is 0.85, and the F1-score for detecting scoliosis (with a Cobb angle greater than 10 degrees) is 0.96, demonstrating the ability to achieve scoliosis detection early.

\section{Discussion}
In traditional daily practice, Cobb angle measurements rely on the readers to visually determine which two vertebrae have the most tiled angle and measure the angle between, and this suffers a great inter-reader and intra-reader variability. When comes to an automatic measurement algorithm, the difficulty in accurately detecting and identifying each individual vertebra would increase the instability of measurements.

In this paper, we propose a fully automatic, end-to-end Cobb angle measurement method for adolescent idiopathic scoliosis (AIS). Instead of relying on detecting each individual vertebra, which can be problematic because of the difficulties of extracting each vertebra, our method measures the Cobb angle by using the geometry of the whole spine based on the interpretation of the definition of the Cobb angle in a mathematical manner.

Our experimental results show our method achieves a mean CA difference of 4.17 $[(CI): 3.54^\circ, 4.79^\circ]$. This deviation is notably lesser than the CA difference of 5.17 $[(CI): 4.31^\circ, 6.04^\circ]$ observed amongst the readers themselves under the p-value of 0.03. Additionally, when juxtaposed with the documented 95\% confidence interval in literature, which ranges between 2.5 to 8.8 degrees for experienced readers \cite{gstoettner2007inter}, our model achieves a more constrained range of difference compared with the readers in the field.

Furthermore, as evidenced by Row 1 in Figure \ref{fig:quantative_result}, which focuses on the Cobb Angle (CA) discrepancies between the algorithm and human readers for each case, cases with elevated levels of divergence also exhibit pronounced CA variance among the readers themselves (as denoted by the orange bars). This observation underscores the inherently challenging nature of these cases, indicating a difficulty among the readers in arriving at a consensus. When we visualized all the cases larger than 5 degrees of differences (row 2, at Figure \ref{fig:quantative_result}), we found our algorithm achieved a consensus of at least one group of readers’ measurements in 8 out of these 10 cases, and only in 2 cases do we have different measurements.

The pairwise comparison with 7 readers including MAD, ICC, Pearson correlation coefficient and Cohen correlation coefficient shown in Figure \ref{fig:MAE_result}, shows that our algorithm can have greater commonalities and agreement with the other readers compared to the readers' degree of agreement with each other. This also reflects, to some extent, the inter-reader stability that our algorithm possesses. It not only achieves the same level of stability as experienced physicians but also compensates to some extent for the bias caused by individual reader measurements, which indicates a great potential to use it in real-world settings to reach a more consistent measurement standard between readers and between cases.
There are certain limitations to this study. The instability issues persist in our algorithm, which is founded upon two-dimensional imaging, as a result of the projection challenges posed by X-rays, which differ depending on the posture of the patient. In clinical settings, however, it is common practice to measure the Cobb angle with 2D X-rays; therefore, we consider our instrument to be more practical for such applications. Additionally, our method of measurement may be marginally distinct from those that rely on the detection of individual vertebrae. However, following an exhaustive comparison with reader evaluations, we are certain that our measurements are adequately consistent with those of the readers.

\section{Conclusion}
In our study, we proposed a three-step automatic Cobb angle measurement algorithm as well as a comprehensive reader study.
The comparison of our algorithm measurements with those of multiple expert readers demonstrated that our approach delivers the highest level of reader consensus and reader-level performance. Also, our algorithm selects the locations to measure based on the overall geometric properties of the spine, which have a high level of interpretability and reproducibility. In the future, our algorithm could be applied to clinical diagnosis to assist doctors with scoliosis assessments and diagnosis tasks. 

\appendix

 \bibliographystyle{elsarticle-num} 
 \bibliography{cas-refs}





\end{document}